\documentclass[10pt,twocolumn,aps,pra,amsmath,amssymb,showpacs]{revtex4-2}
\usepackage{bm}
\usepackage{mathrsfs}
\usepackage{graphicx}
\usepackage[usenames,dvipsnames,svgnames,table]{xcolor}
\usepackage{framed}
\usepackage{amsthm}
\usepackage[utf8]{inputenc}

\colorlet{shadecolor}{blue!15}
\usepackage[unicode=true,
            pdfusetitle, 
            bookmarks=true,
            bookmarksnumbered=false,
            bookmarksopen=false,
            breaklinks=true,
            pdfborder={0 0 0},
            backref=false,
            colorlinks=true]{hyperref}
\hypersetup{linkcolor=NavyBlue,urlcolor=NavyBlue,citecolor=NavyBlue}

\usepackage{amsthm}
\newcommand{\propnumber}{} 
\newtheorem*{prop}{Proposition \propnumber}

  \newtheorem{propo}{Proposition} 

\newtheorem{lemma}{Lemma}

 \newtheorem{cor} {Corollary}
 \newtheorem{rem}{Remark}

 \usepackage{tikz-cd}

\newcommand{\cH}{\mathcal{H}}

\begin{document}

\title {Instantaneous measurement can isolate the information}

\author {Iman Sargolzahi}
\email{sargolzahi@um.ac.ir}

\affiliation{Department of Physics, Faculty of Science, Ferdowsi University of Mashhad, Mashhad, Iran}

\begin{abstract}
Consider a one-dimensional spin chain, from spin $1$ to spin $N$, such that each spin interacts  with its nearest neighbors. 
Performing a local operation (measurement) on spin $N$, we expect from the Lieb-Robinson velocity that, in general, the effect of this measurement achieves  spin $1$ after some while. 
But, in this paper, we show that if a)  the measurement on spin $N$ is performed instantaneously and b) the initial state of the  spin chain is chosen appropriately, then the effect of the measurement on spin  $N$ never achieves spin $1$.
In other words, performing or not performing an instantaneous measurement on spin $N$ at  $t=0$ does not alter the reduced dynamics of spin $1$ for  all the times $t\geq 0$. We can interpret this as the following: The information of performing an instantaneous measurement on spin $N$ is isolated such that it cannot achieve spin $1$.
\end{abstract}

\maketitle

\section{Introduction}


Consider two spins (qubits) $1$ and $N$, which  are initially prepared in a state $\rho_{1N}$ and then are separated far from each other such that they cannot interact. Now, performing a local measurement on spin $N$ changes the initial state $\rho_{1N}$ to
\begin{equation}
\label{eq:1}
\begin{aligned}
 \rho_{1N}^\prime= \mathrm{id}_1\otimes\mathcal{M}_N ( \rho_{1N}),
\end{aligned}
\end{equation}
where $\mathrm{id}_1$ is the identity map on spin $1$, and $\mathcal{M}_N$ is a trace-preserving (completely positive) map on spin $N$ \cite{1}. So, the reduced state of  spin $1$ after the measurement, i.e., 
$\rho_1^\prime = \mathrm{Tr}_{N}(\rho_{1N}^{\prime})$, is the same as its initial state $\rho_1 = \mathrm{Tr}_{N}(\rho_{1N})$.

This result that $\rho_1^\prime=\rho_1$, regardless of whether the initial $\rho_{1N}$ is correlated or not, is interpreted as the following: No information of performing the measurement $\mathcal{M}_N$ on spin $N$ can achieve spin $1$. Therefore, quantum mechanics and special relativity are not inconsistent, even if the measurement $\mathcal{M}_N$ is performed instantaneously, and so the whole state of spins $1$ and $N$ changes instantaneously from $ \rho_{1N}$ to $\rho_{1N}^\prime$.

This fact that the reduced state of spin $1$ does not change due to the measurement  performed on spin $N$ is not such unexpected, since spins $1$ and $N$ are far from each other and  do not interact. But,  if spins $1$ and $N$ are connected to each other by a spin chain, from spin $2$ to spin $N-1$,  we  expect that the information about performing the measurement $\mathcal{M}_N $ on spin $N$  achieves spin $1$ after some while.

Let us consider the simple case that each spin  $j$, in the spin chain from $1$ to $N$, interacts only with its nearest neighbors, and with an external local magnetic field $b_j$ which is along the $x$-axis. So, the Hamiltonian is
\begin{equation}
\label{eq:2}
\begin{aligned}
H=\sum_{j=1}^{N-1} Z_j \otimes Z_{j+1} + \sum_{j=1}^N b_j X_j , 
\end{aligned}
\end{equation}
where $ X_j$ and  $ Z_j$ are the first and the third Pauli operators on spin $j$. Assume that all  $b_j$ are zero, except  $b_N$ which is turned on at  $t=0$. Now, we ask whether  spin $1$ becomes aware of the existence of $b_N$, or not. If the term $b_N X_N$  commuted with the rest of the Hamiltonian in Eq. \eqref{eq:2}, then the reduced state of the other spins, including spin $1$, would  not change after turning on the  $b_N$, but now, we expect that spin $1$ will become aware of  $b_N$ after some while.

Similar line of reasoning can be given for any localized quantum operation $\mathcal{F}_N$, which is performed on spin $N$ during a time interval $[t_1, t_2]$. If $\mathcal{F}_N$ commutes with the   time evolution operator
  $U=\mathrm{exp}(-iHt)$, where $i=\sqrt{-1}$, $t$ is the time and $H$ is the  Hamiltonian in Eq \eqref{eq:2},
  then the reduced state of spin $j \neq N$ (including spin $1$) does not change. Otherwise, in general, we expect that the information of performing $\mathcal{F}_N$ on spin $N$ will achieve spin $1$. We will  discuss this case (and also the case mentioned in the previous paragraph) in more detail in the two following sections.

What does happen if the local operation on spin $N$ is performed instantaneously? 
In the standard text-book quantum mechanics, there exists one operation which  is performed instantaneously, i.e., the  measurement. In Sec. \ref{sec: B2},
using the Lieb-Robinson bound \cite{b1, b2}, we show that regardless of whether the measurement on spin $N$ is 
instantaneous or not, in general, its influence will achieve spin $1$.

The main question of this paper is whether it is possible to find an operation (measurement) on spin $N$ such that its influence never achieves spin $1$.
 We show that the answer is affirmative, at least, if
  we perform an instantaneous measurement $\mathcal{M}_N$ on spin $N$,  and if we choose the initial state of the whole spin chain from $1$ to $N$ appropriately. This result is given in Sec. \ref{sec: D}, after giving some  preliminaries needed in  Sec. \ref{sec: C}.

When the      measurement is not  instantaneous, 
 whether our results in Sec.  \ref{sec: D} can be used 
 is discussed in Sec. \ref{sec: E}.  Finally, we end this paper in  Sec. \ref{sec: F}, with a summary of our results.

\section{Locality in a spin chain: when the measurement operation commutes with the free evolution}  \label{sec: B1} 

Consider a spin chain, from spin $1$ to spin $N$, such that each spin interacts with its nearest neighbors through the Hamiltonian 
\begin{equation}
\label{eq:b1}
\begin{aligned}
H=\sum_{j=1}^{N-1} Z_j \otimes Z_{j+1}.
\end{aligned}
\end{equation}
Then, we add a local term $H_N(t)$ to the Hamiltonian such that this (in general) time-dependent term acts only on spin $N$ and $[H, H_N(t)]=0$.
We may assume that $H_N(t)$ describes the interaction of spin $N$ with some external field which is turned on  at $t=0$. Other spins, including spin $1$, will never become aware of  existence of this new term  $H_N(t)$ in the total Hamiltonian $H_T=H+H_N(t)$. This  can be proved simply, using the interaction picture.

Assume that the state of the whole spin chain at time $t$ is given by $\rho_\Lambda(t)$, in the Schrodinger picture. ($\Lambda$ denotes the whole spin chain.)  We define the state of the  spin chain, at this $t$, in the interaction picture as
\begin{equation}
\label{eq:b2}
\begin{aligned}
\rho^I_\Lambda(t)=U^{\dagger} \rho_\Lambda(t) U,
\end{aligned}
\end{equation}
where $U=\mathrm{exp}(-iHt)$ with the free Hamiltonian in Eq. \eqref{eq:b1}. So, the time evolution of the density operator in the interaction picture is given by 
\begin{equation}
\label{eq:b3}
\begin{aligned}
\partial_t\rho^I_\Lambda(t)=-i [H_N(t),\rho^I_\Lambda(t)],
\end{aligned}
\end{equation}
where we have used this fact that $H_N(t)$ commutes with $U$. Therefore, 
\begin{equation}
\label{eq:b4}
\begin{aligned}
\rho^I_\Lambda(t)=\tilde{U}_N(t)\rho^I_\Lambda(0) \tilde{U}_N^\dagger(t)  \\
\qquad=\tilde{U}_N(t)\rho_\Lambda(0) \tilde{U}_N^\dagger(t), 
\end{aligned}
\end{equation}
where $\tilde{U}_N(t)$ is a unitary operator which is a function of $H_N(t^\prime)$ from $t^\prime=0$ to $t^\prime=t$, and so, acts only on spin $N$.  Using Eqs.  \eqref{eq:b2} and \eqref{eq:b4}, we have
\begin{equation}
\label{eq:b5}
\begin{aligned}
\rho_\Lambda(t)=U\tilde{U}_N(t)\rho_\Lambda(0) \tilde{U}_N^\dagger(t)U^\dagger  \\
\qquad=\tilde{U}_N(t)U\rho_\Lambda(0)U^\dagger \tilde{U}_N^\dagger(t), 
\end{aligned}
\end{equation}
where we have used this fact that $[U, \tilde{U}_N(t)]=0$. Finally, the reduced state of the whole spin chain except spin $N$, which we  denote as $\Lambda \backslash N$, is given by 
\begin{equation}
\label{eq:b6}
\begin{aligned}
\rho_{\Lambda\backslash N}(t)=\mathrm{Tr}_{N}(\rho_\Lambda(t))=\mathrm{Tr}_{N}(U\rho_\Lambda(0)U^\dagger).
\end{aligned}
\end{equation}
So, the reduced state of  other spins will not be affected by $\tilde{U}_N(t)$. In other words, other spins, even spin $N-1$ which interacts with its neighbor spin $N$ through $H$ in Eq.  \eqref{eq:b1}, will never become aware of the interaction between spin $N$ and some external field which causes $H_N(t)$.

We can follow a similar line of reasoning when spin $N$ undergoes some local operation (measurement) $\mathcal{F}_N$ during a time interval $[t_1, t_2]$. The most general quantum operation $\mathcal{F}_N$ on spin $N$ can be modeled as the interaction of spin $N$ with another quantum  system, which we  call it the probe $P$, through a unitary evolution $U_{NP}$, and then tracing over the probe $P$. In addition, we assume that $U_{NP}$ is due to an interaction Hamiltonian $H_{NP}(t)$, which is performed  on $N$ and $P$ during the time interval $[t_1, t_2]$. Mathematically, one can always find such an interaction Hamiltonian $H_{NP}(t)$, assuming that the probe $P$ is only an ancillary system \cite{1}. But, (at least) when $\mathcal{F}_N$ describes a real measurement on spin $N$, we would like to consider the  probe $P$ as a real physical system, and so $H_{NP}(t)$ as the real interaction Hamiltonian between spin $N$ and the probe $P$. (See also Sec. \ref{sec: E}.)

 Let us denote the state of the whole spin chain $\Lambda$ and the probe $P$ (in the Schrodinger picture) as $\rho_{\Lambda P}(t)$. We  assume that before  beginning of the measurement, the spin chain (spin $N$) and the probe do not interact with each other.
  So, before  beginning of the measurement, i.e., for all the times $t^\prime \leq t_1$ ($t_1 \geq 0$), we have 
 \begin{equation}
\label{eq:b7}
\begin{aligned}
\rho_{\Lambda P}(t^\prime)=\rho_{\Lambda }(t^\prime)\otimes \rho_{P}(t^\prime),
\end{aligned}
\end{equation}
where $\rho_{\Lambda }(t^\prime)$ and $\rho_{P}(t^\prime)$ are the states of the spin chain and the probe at  time $t^\prime$, respectively. 
 
 Then, we define the state of spin chain and the probe, in the interaction picture, as
\begin{equation}
\label{eq:b8}
\begin{aligned}
\rho^I_{\Lambda P}(t)=U^{\dagger}\otimes I_P \rho_{\Lambda P}(t) U\otimes I_P,
\end{aligned}
\end{equation}
where, as before, $U=\mathrm{exp}(-iHt)$ with the free Hamiltonian in Eq. \eqref{eq:b1}, and $I_P$ is the identity operator on $P$. One can replace $I_P$ with another unitary $U_P$ which describes the free evolution of the probe $P$. But, we simply assume that the state of the probe does not change, unless it interacts with  spin $N$.

Using Eq. \eqref{eq:b8}, we have
\begin{equation}
\label{eq:b9}
\begin{aligned}
\partial_t\rho^I_{\Lambda P}(t)=-i [H_{NP}(t),\rho^I_{\Lambda P}(t)],
\end{aligned}
\end{equation}
where, as before,  we have assumed that $[H_{NP}(t), H]=0$.
 So, 
\begin{equation}
\label{eq:b10}
\begin{aligned}
\rho^I_{\Lambda P}(t)=\tilde{U}_{NP}(t)\rho^I_{\Lambda P}(0) \tilde{U}_{NP}^{\dagger} (t) \qquad\quad \ \\
\qquad=\tilde{U}_{NP}(t)\rho_\Lambda(0)\otimes \rho_P(0) \tilde{U}_{NP}^{\dagger} (t), 
\end{aligned}
\end{equation}
where $\tilde{U}_{NP}(t)$ is a unitary operator which acts only on spin $N$ and the probe $P$,  since it is a function of $H_{NP}(t^\prime)$ from $t^\prime=t_1$ to $t^\prime=t_2$. (For $t^\prime \in [0, t_1)$ and  $t^\prime \in (t_2, t]$, the measurement interaction $H_{NP}(t^\prime)$ is zero, by assumption.)

Using Eqs. \eqref{eq:b8} and \eqref{eq:b10}, we have
\begin{equation}
\label{eq:b11}
\begin{aligned}
\rho_{\Lambda P}(t)=U\tilde{U}_{NP}(t)\rho_\Lambda(0)\otimes \rho_P(0) \tilde{U}_{NP}^{\dagger}(t)U^\dagger \ \\
 :=\mathcal{U} \circ \tilde{\mathcal{U}}_{NP}(t) [\rho_\Lambda(0)\otimes \rho_P(0)]. \qquad \quad
\end{aligned}
\end{equation}
Tracing over the probe $P$, we get
\begin{equation}
\label{eq:b12}
\begin{aligned}
\rho_{\Lambda}(t)=\mathrm{Tr}_{P}(\rho_{\Lambda P}(t)) \qquad\qquad\qquad\qquad\quad  \\ 
=\mathcal{U} \circ \mathrm{Tr}_{P} \circ \tilde{\mathcal{U}}_{NP}(t) [\rho_\Lambda(0)\otimes \rho_P(0)] \ \\
:= \mathcal{U} \circ \mathcal{F}_{N} [\rho_\Lambda(0)], \qquad\qquad\qquad\qquad
\end{aligned}
\end{equation}
where $\mathcal{F}_{N} [*]= \mathrm{Tr}_{P} \circ \tilde{\mathcal{U}}_{NP}(t) [*\otimes \rho_P(0)]$ is a quantum operation, i.e., a trace-preserving completely positive map, on spin $N$. 
In addition, this fact that $[H_{NP}(t), H]=0$ results that, in Eq. \eqref{eq:b11},  $\mathcal{U}$ and  $\tilde{\mathcal{U}}_{NP}(t)$  commute. So, $\rho_{\Lambda P}(t)=\tilde{\mathcal{U}}_{NP}(t) \circ \mathcal{U}\ [\rho_\Lambda(0)\otimes \rho_P(0)]$. Therefore,
\begin{equation}
\label{eq:b13}
\begin{aligned}
\rho_{\Lambda}(t)=\mathrm{Tr}_{P}(\rho_{\Lambda P}(t))
=   \mathcal{F}_{N} \circ \mathcal{U} \ [\rho_\Lambda(0)].
\end{aligned}
\end{equation}
Comparing Eqs. \eqref{eq:b12} and \eqref{eq:b13} shows that quantum operations $ \mathcal{F}_{N}$ and $\mathcal{U}$ commute. Note that, although the measurement $ \mathcal{F}_{N}$ is performed during the time interval $[t_1, t_2]$ which is within the time interval $[0, t]$, from  Eqs. \eqref{eq:b12} and \eqref{eq:b13},  we see that it  can be considered   as (an instantaneous) measurement  performed (right) before or after the unitary time evolution $\mathcal{U}$, respectively. ($\mathcal{U}$ is performed during the time interval $[0, t]$.)

Tracing over spin $N$ in Eq.  \eqref{eq:b13}, we conclude that
\begin{equation}
\label{eq:b14}
\begin{aligned}
\rho_{\Lambda \backslash N}(t)=\mathrm{Tr}_{N}(\rho_{\Lambda}(t))
=  \mathrm{Tr}_{N} \circ \mathcal{U} \ [\rho_\Lambda(0)],
\end{aligned}
\end{equation}
since  $ \mathcal{F}_{N}$ is a trace-preserving operation on spin $N$. Therefore, other spins will not become aware of performing the measurement  $ \mathcal{F}_{N}$ on spin $N$.

Let us summarize the results of  this section as the following:
\begin{cor}
\label{cor1}
Consider a  spin chain  with the free Hamiltonian $H$ in Eq. \eqref{eq:b1}. If the interaction Hamiltonian, $H_N(t)$ or $H_{NP}(t)$, commutes with  $H$, other spins  never become aware of performing the local operation on spin $N$, generated by $H_N(t)$ or $H_{NP}(t)$. 
\end{cor}
  But, when the interaction Hamiltonian does not commute with $H$, we expect that, in general, the reduced state of other spins will be affected. We  show this in the next section, using the Lieb-Robinson bound \cite{b1, b2, b3}.
  
\section{Locality in a spin chain: when the measurement operation does not commute with the free evolution}  \label{sec: B2}

First, we consider the case of non-instantaneous measurement governed by $H_{NP}(t)$.
So, the total Hamiltonian is
\begin{equation}
\label{eq:b15}
\begin{aligned}
H_T(t)=H+H_{NP}(t),
\end{aligned}
\end{equation}
where $H$ is given in Eq. \eqref{eq:b1}, and $[H, H_{NP}(t)] \neq 0$. The case mentioned in the Introduction, for which the total Hamiltonian is given in Eq. \eqref{eq:2},  where only $H_N=b_n X_N$ is nonzero, can be treated similarly.

The total Hamiltonian  in Eq. \eqref{eq:b15} is an example of finite range Hamiltonians, since each site
interacts only with its nearest neighbors. (We have considered the probe $P$ as a neighbor of spin $N$.) 
For such kind of  Hamiltonians,  the Lieb-Robinson bounds, as upper bounds on the norm of the commutators of local operators  at different times, have been proved \cite{b1, b2}.

For example, assume that $O_1$ and $O_N$ are Hermitian operators (observables in the Schrodinger picture) acting on spin $1$ and spin $N$, respectively. In the Heisenberg picture, an observable evolves as 
\begin{equation}
\label{eq:b16}
\begin{aligned}
O^H(t)=U_T^\dagger(t) O U_T(t),
\end{aligned}
\end{equation}
where $U_T(t)$ is the unitary evolution operator generated by the total Hamiltonian $H_T$.
Now, for the local observables  $O_1$ and $O_N$, we have the following  Lieb-Robinson bound \cite{b1}:
\begin{equation}
\label{eq:b17}
\begin{aligned}
\parallel [O_1^H(t), O_N]\parallel \leq c \parallel O_1\parallel  \parallel O_N\parallel \exp(-\frac{L-vt}{\xi}),
\end{aligned}
\end{equation}
where $L=N-1$ is the distance between spin $1$ and spin $N$, and  $\parallel O\parallel$ is the norm of the operator $O$. When $O$ is  Hermitian,  its norm   is  the absolute value of its largest eigenvalue \cite{b4}. In addition, $c$, $\xi$ and $v$ are positive constants, depending on the norm of different terms in the total Hamiltonian $H_T$ in Eq. \eqref{eq:b15}. They also depend on  the range of different terms in $H_T$ . (Different terms of $H_T$ in Eq. \eqref{eq:b15} have the same range $2$, i.e., each term connects two sites to each other.)   We can call $v$ as  the Lieb-Robinson velocity, which limits the speed of  propagation of information in the spin chain.

We define the set $B_l$ as the set of all sites  that their distance from spin $1$ is at most the integer $l$.
The restriction of $O_1^H(t)$ to this subspace is 
\begin{equation}
\label{eq:b18}
\begin{aligned}
O_1^{H_l}(t)=: \mathrm{Tr}_{(\Lambda P) \backslash B_l}(O_1^H(t))\otimes \frac{I_{(\Lambda P)\backslash B_l}}{\mathrm{Tr}(I_{(\Lambda P)\backslash B_l})}, 
\end{aligned}
\end{equation}
where $I_{(\Lambda P)\backslash B_l}$ is the identity operator on the subspace $(\Lambda P) \backslash B_l$, i.e., all the sites outside the ball $B_l$. Now, using the  Lieb-Robinson bound in Eq. \eqref{eq:b17}, it can be shown that \cite{b3}:
\begin{equation}
\label{eq:b19}
\begin{aligned}
\parallel O_1^H(t) - O_1^{H_l}(t)\parallel \leq c \parallel O_1\parallel \exp(-\frac{l-vt}{\xi}).
\end{aligned}
\end{equation}
So, (only) when $l$, i.e., the radius of the ball $B_l$, is large compared to $vt$, $O_1^{H_l}(t)$ is a good approximation of $ O_1^H(t)$. In other words, the operator  $ O_1^H(t)$ which was first, i.e., at $t=0$, localized at spin $1$, gradually spreads over other sites, due to the neighbors interaction through $H_T$. The speed of this spread is restricted by  the Lieb-Robinson velocity $v$.

We can use this result to show   how the local term $H_{NP}$ in Eq. \eqref{eq:b15}, i.e., the interaction between the probe $P$ and the spin $N$, can affect the evolution of $ O_1^H(t)$, which is an observable on spin $1$.

First, we write the  Heisenberg evolution equation for the operator $ O_1^H(t)$. Using Eq. \eqref{eq:b16}, we have
\begin{equation}
\label{eq:b20}
\begin{aligned}
\partial_t O_1^H(t) = i [H_T^H(t), O_1^H(t)],
\end{aligned}
\end{equation}
where $H_T^H(t)$ is also defined as Eq. \eqref{eq:b16}. The total Hamiltonian in Eq. \eqref{eq:b15} includes only local terms, i.e., each term in $H_T$ acts only on two neighbor sites. So, for each term in $H_T$, a relation similar to Eq. \eqref{eq:b19} also holds. In other words, each first local term in $H_T^H$ grows over other sites with the speed (limited to) $v$.


From Eq. \eqref{eq:b20}, we see that only those terms of $H_T^H(t) $ which do not commute with $O_1^H(t)$ affect the evolution of  $O_1^H(t)$. So, we expect that (at least) from an instant $\tilde{t}$, for which  $H_{NP}^H(\tilde{t})$ starts to overlap  with   $O_1^H(\tilde{t})$, the local observable  $O_1$ will be affected by $H_{NP}$.
Since the supports of $H^H_{NP}(t)$ and $O_1^H(t)$ grow with the speed $v$, we expect  that  $H_{NP}$ will affect the local observables on spin $1$ for  times $t>t_1 +\frac{L}{2v}$. ($H_{NP}(t)$  is turned on at $t=t_1$.)

Therefore, when the measurement on spin $N$ is governed (modeled) by the interaction Hamiltonian  $H_{NP}(t)$ which lasts from $t_1$ to $t_2$  and $[H, H_{NP}(t)] \neq 0$,  spin $1$, in general, will become aware of this measurement after some while.

 A similar result can be proved for instantaneous measurements too, following a similar procedure as what given in Ref. \cite{b3}.

Assume that the initial state of the spin chain is $\rho_\Lambda (0)$, and the initial state of (an ancillary) probe is $\rho_P (0)$. An arbitrary instantaneous measurement performed on spin $N$ at $t=0$ can be modeled as implementing a unitary operator $U_{NP}$ on spin $N$ and the probe $P$, and then tracing over the probe $P$. If, after the measurement, the time evolution operator of the spin chain, from $0$ to $t$, is given by $U_\Lambda (t)$, then the final state of the spin chain is
\begin{equation}
\label{eq:b21}
\begin{aligned}
\sigma_\Lambda (t)= \mathcal{U}_\Lambda (t) \circ \mathrm{Tr}_{P} \circ \mathcal{U}_{NP}
  [\rho_\Lambda(0)\otimes \rho_P(0)]  \  \\
=\mathrm{Tr}_{P} \circ 
\mathcal{U}_\Lambda (t) \circ \mathcal{U}_{NP}
  [\rho_\Lambda(0)\otimes \rho_P(0)].
\end{aligned}
\end{equation}
In addition, note that if the measurement on spin $N$ is not implemented, the state of the spin chain at time $t$ can be written as
\begin{equation}
\label{eq:b22}
\begin{aligned}
\rho_\Lambda (t)= \mathcal{U}_\Lambda (t) \circ \mathrm{Tr}_{P} 
  [\rho_\Lambda(0)\otimes \rho_P(0)]  \  \\
=\mathrm{Tr}_{P} \circ 
\mathcal{U}_\Lambda (t) 
  [\rho_\Lambda(0)\otimes \rho_P(0)].
\end{aligned}
\end{equation}

We want to compare between the expectation values of an observable $O_1$ on spin $1$, performing or not performing an instantaneous measurement on spin $N$ at $t=0$. The absolute value of the difference between these two expectation values at time $t$ is
\begin{equation}
\label{eq:b23}
\begin{aligned}
\mid \mathrm{Tr} (O_1 (\rho_\Lambda (t)-\sigma_\Lambda (t))\mid  \qquad \qquad \qquad \qquad\qquad    \qquad\quad \ \\
=\mid \mathrm{Tr} (O_1^H(t)\rho_\Lambda(0)\otimes \rho_P(0)) \qquad \qquad\qquad\qquad  \  \\  
-\mathrm{Tr} (O_1^H(t)U_{NP}\rho_\Lambda(0)\otimes \rho_P(0) U_{NP}^{\dagger})\mid \\
=\mid \mathrm{Tr} (\rho_\Lambda(0)\otimes \rho_P(0)  U_{NP}^{\dagger} [U_{NP}, O_1^H(t)])\mid \qquad \  \  \\
\leq \parallel  U_{NP}^{\dagger} [U_{NP}, O_1^H(t)]\parallel \qquad \qquad \qquad \qquad \quad \quad \\
= \parallel  [U_{NP}, O_1^H(t)]\parallel,    \qquad \qquad \qquad \qquad\qquad \qquad 
\end{aligned}
\end{equation}
where $O_1^H(t)=U_{\Lambda}^\dagger(t) O_1 U_{\Lambda}(t)$. In going from  line four to  line five, we have used this fact that the absolute value of the expectation value of the Hermitian operator $A= U_{NP}^{\dagger} [U_{NP}, O_1^H(t)]$ is less than or equal to its norm.
In addition,  in going from  line five to  line six, we have used this fact that the norm of the linear operator $U_{NP}A$ is equal to the norm of $A$ \cite{b4}.

Using Eqs.  \eqref{eq:b17} and  \eqref{eq:b23}, we conclude that
\begin{equation}
\label{eq:b24}
\begin{aligned}
\mid \mathrm{Tr} (O_1 (\rho_\Lambda (t)-\sigma_\Lambda (t))\mid  
\leq  \parallel  [U_{NP}, O_1^H(t)]\parallel \qquad\quad   \\
 \leq c \parallel O_1\parallel  \exp(-\frac{L-vt}{\xi}),   
\end{aligned}
\end{equation}
where we have used this fact that $\vert\vert U_{NP} \vert\vert=1$.
Therefore, for $t \geq \frac{L}{v}$, the difference between the two expectation values $ \mathrm{Tr} (O_1 (\rho_\Lambda (t))$ and $ \mathrm{Tr} (O_1 (\sigma_\Lambda (t))$ is not neglectable, in general. In other words, we expect that performing an instantaneous measurement on spin $N$, its influence will  achieve spin $1$ after some while.


Let us summarize the main result of this section as the following:
\begin{cor}
\label{cor2}
Unless the case mentioned in  Corollary \ref{cor1}, we expect that, in general, the influence of a local measurement on spin $N$ achieves spin $1$  after a period of time,
 regardless of whether the measurement on spin $N$ is instantaneous or not.
\end{cor}

Note that the Lieb-Robinson bound in Eq. \eqref{eq:b17}, and so Eqs. \eqref{eq:b19} and \eqref{eq:b24}, give us upper bounds. So, to be precise, we should say that  they give us the optimal possible velocity of the information propagation, and so only an estimate of the real such velocity. In fact, the main result of this paper is that one can find cases for which the real velocity of  information propagation is zero. 
Namely, in Sec.  \ref{sec: D}, we find cases for which we have $ \mathrm{Tr} (O_1 (\rho_\Lambda (t))= \mathrm{Tr} (O_1 (\sigma_\Lambda (t))$, for arbitrary observable $O_1$ on spin $1$  and for any time $t \geq 0$. Therefore, the left hand side of inequality  \eqref{eq:b24} is always zero for these cases.

We end this section, with the following point.
Assume that $U_{NP}$ in Eq. \eqref{eq:b21} depends on some real parameter $\theta$ as $U_{NP}(\theta)=\exp(-i\theta \tilde{H}_{NP})$, where $\tilde{H}_{NP}$ is a local Hermitian operator  on spin $N$ and the (ancillary) probe $P$.
  So, in general, the final state of spin $1$ also depends on  $\theta$:
\begin{equation}
\label{eq:b25}
\begin{aligned}
\sigma_1^{(\theta)} (t)
=\mathrm{Tr}_{(\Lambda P) \backslash 1} \circ 
\mathcal{U}_\Lambda (t) \circ \mathcal{U}_{NP} (\theta)
  [\rho_\Lambda(0)\otimes \rho_P(0)].
\end{aligned}
\end{equation}
Consider a local POVM $\lbrace E_\gamma \rbrace$ on spin $1$. This means that $ E_\gamma$ are positive operators on spin $1$ such that $\sum_\gamma E_{\gamma}=I_1$, where $I_1$ is the identity operator on spin $1$. So, $\lbrace E_\gamma \rbrace$ describes a measurement on spin $1$, such that the probability of the outcome $\gamma$ is given by $P(\gamma \vert \theta)=\mathrm{Tr}[E_\gamma \sigma_1^{(\theta)} (t)]$ \cite{1, b7}. 
The Fisher information for this measurement is defined as:
\begin{equation}
\label{eq:b26}
\begin{aligned}
F(\theta)=\sum_\gamma \frac{1}{P(\gamma \vert \theta)} [\partial_\theta P(\gamma \vert \theta)]^2,
\end{aligned}
\end{equation}
which is a measure of the sensitivity of  the measurement  $\lbrace E_\gamma \rbrace$ to  small changes in $\theta$ \cite{b6, b7}.

The maximum of the Fisher information, over all possible  POVMs, is called the quantum Fisher
information and is given by \cite{b6, b7}:
\begin{equation}
\label{eq:b27}
\begin{aligned}
F_Q [\sigma_1^{(\theta)} (t)]=\sum_{j,k=1}^2 \frac{2}{p_j+p_k}  \vert \langle \psi_j \vert \partial_\theta
\sigma_1^{(\theta)} (t) \vert \psi_k \rangle \vert^2, 
\end{aligned}
\end{equation}
where $p_{j\slash k}$ and $\vert \psi_{j\slash k} \rangle$ are the eigenvalues and eigenvectors of  $\sigma_1^{(\theta)} (t)$, respectively. (The summation is over those $j$ and $k$ for which $ p_j+p_k > 0$.)

Using the  Lieb-Robinson bound in Eq. \eqref{eq:b17}, it has been shown in Ref. \cite{b5} that
\begin{equation}
\label{eq:b28}
\begin{aligned}
F_Q [\sigma_1^{(\theta)} (t)] \leq a(t) \exp(-\frac{L-vt}{\xi}),
\end{aligned}
\end{equation}
where the positive coefficient $a(t)$ depends on the constant $c$ in Eq. \eqref{eq:b17}, and also on $\sigma_1^{(\theta)} (t)$ and $\tilde{H}_{NP}$. Therefore, when $vt \ll L$, the  sensitivity of  the measurements on spin $1$ to (small changes in) $\theta$ is neglectable.
In other words,  in general, the information of performing $U_{NP}(\theta)$ on spin $N$ and the probe $P$ propagates in the spin chain with the  Lieb-Robinson velocity $v$.

\section{Identical reduced dynamics}  \label{sec: C}

As we have seen, the  Lieb-Robinson velocity $v$ gives us an upper bound on the speed of  propagation of the  influence (information) of a local measurement on spin $N$.
Now, we ask whether there exists a lower bound for the speed of propagation of  information in the spin chain, or not.
In the next section, we show that the optimal probable lower bound, i.e., the zero speed, is in fact possible, at least, when the  measurement on spin $N$ is instantaneous. Before, we need some preliminaries which are given in the following.

Consider a quantum system $S$, interacting with its environment $E$. The whole system-environment is a closed quantum system, which evolves  unitarily as \cite{1}:
\begin{equation}
\label{eq:3}
 \rho_{SE}^{\prime}=  U \rho_{SE} U^{\dagger},
 \end{equation} 
 where $U$ is a unitary operator, on $\cH_S\otimes\cH_E$. $\cH_S$ and $\cH_E$ are the Hilbert spaces of the system and the environment, respectively.  In addition, $\rho_{SE}$ and $\rho_{SE}^{\prime}$ are  initial and  final states (density operators) of the system-environment, respectively.
So, the reduced dynamics of the system is given by  
\begin{equation}
\label{eq:4}
\rho_{S}^{\prime}=\mathrm{Tr}_{E}(\rho_{SE}^{\prime})=\mathrm{Tr}_{E} (U \rho_{SE} U^{\dagger}).
\end{equation}

Consider another initial state of the system-environment $\sigma_{SE}$ such that $\sigma_{S}=\mathrm{Tr}_{E}(\sigma_{SE})=\mathrm{Tr}_{E}(\rho_{SE})=\rho_S$, i.e., though $\sigma_{SE}$ differs from 
$\rho_{SE}$, but the initial state of the system  is the same for both $\rho_{SE}$ and $\sigma_{SE}$.  
So,
 \begin{equation}
\label{eq:5}
\begin{aligned}
\sigma_{SE}= \rho_{SE}+ R,
\end{aligned}
\end{equation}
where $R$ is a Hermitian operator, on $\cH_S\otimes\cH_E$, such that $\mathrm{Tr}_{E}(R)=0$.
Now, we ask when the final state of the system is also the same, for both initial states of the system-environment $\rho_{SE}$ and $\sigma_{SE}$. In other words, when is $\sigma_{S}^{\prime}=\mathrm{Tr}_{E} (U \sigma_{SE} U^{\dagger})$ the same as $\rho_{S}^{\prime}$ in Eq. \eqref{eq:4}? A similar question is addressed in Refs. \cite{3, 4, 5}.

In the following, we consider the case that the unitary time evolution of the whole  system-environment $U$ is as $U=\exp (-iHt)$, where $H$ is a time-independent Hamiltonian (a Hermitian operator on $\cH_S\otimes\cH_E$).
In order to $\sigma_{S}^{\prime}=\rho_{S}^{\prime}$, we must have
\begin{equation}
\label{eq:6}
\begin{aligned}
\mathrm{Tr}_{E} (U R U^{\dagger})=0.
\end{aligned}
\end{equation}
Using the Baker-Hausdorff formula
\begin{equation*}
\begin{aligned}
e^{(-T)} Re^T= R+ [R,T]+\frac{1}{2!}[[R,T],T]  \\ +  \frac{1}{3!}[[[R,T],T],T]+ \dots ,
\end{aligned}
\end{equation*}
where $T=iHt$, and Eq. \eqref{eq:6}, we have
\begin{equation}
\label{eq:7}
\begin{aligned}
\mathrm{Tr}_{E}(R)+ \mathrm{Tr}_{E}([R,T])+\frac{1}{2!}\mathrm{Tr}_{E}([[R,T],T]) + \dots=0.
\end{aligned}
\end{equation}
In order that the above relation be valid for arbitrary $t$, the coefficients of the different powers of $t$ must be zero, simultaneously. So, we must have $\mathrm{Tr}_{E}(A_k)=0$, where
\begin{equation}
\label{eq:8}
\begin{aligned}
A_0=R, \quad A_1=[R,H], \quad  A_2=[[R,H],H], \quad \dots .
\end{aligned}
\end{equation}

Now, we expand the Hamiltonian $H$ as 
\begin{equation}
\label{eq:9}
\begin{aligned}
H= H_S \otimes I_E + I_S \otimes H_E + H_{SE},
\end{aligned}
\end{equation}
where $H_S$ and $H_E$ are Hermitian operators, on $\cH_S$ and $\cH_E$, respectively. Also, $I_S$ and $I_E$ are identity operators, on $\cH_S$ and $\cH_E$, respectively. Obviously, the Hermitian operator $H_{SE}$ is $H- H_S \otimes I_E - I_S \otimes H_E$, and we can call it the interaction term of the  Hamiltonian.
\begin{lemma}
\label{lem1}
Consider a linear operator $A_k$, on $\cH_S\otimes \cH_E$, for which we have $\mathrm{Tr}_{E}(A_k)=0$. Then, we also have $\mathrm{Tr}_{E}(A_k H_S \otimes I_E)=0$, $\mathrm{Tr}_{E}(H_S \otimes I_E A_k)=0$ and $\mathrm{Tr}_{E}([A_k, I_S \otimes H_E])=0$.
\end{lemma}
\textit{Proof.}
Assume that the environment $E$ is  finite dimensional, with dimension $d_E$. 
 So, each linear operator $O_E$ on $\cH_E$ can be expanded as
\begin{equation}
\label{eq:10}
\begin{aligned}
O_E=\sum_{\mu=0}^{d_E^2-1} c_\mu G_E^{(\mu)},
\end{aligned}
\end{equation}
where $c_\mu$ are complex coefficients, $G_E^{(0)}=I_E$, and  other $ G_E^{(\mu)}$ are Hermitian traceless operators on $\cH_E$. In addition, all $ G_E^{(\mu)}$  are orthogonal  to each other, with respect to the Hilbert-Schmidt inner product (see, e.g., Ref.  \cite{6}).
Therefore, the linear operator $A_K$, on $\cH_S \otimes \cH_E $, can be expanded as
\begin{equation}
\label{eq:11}
\begin{aligned}
A_k=\sum_{\mu=0}^{d_E^2-1} C_S^{(\mu)}\otimes G_E^{(\mu)},
\end{aligned}
\end{equation}
where $C_S^{(\mu)}$ are linear operators on $\cH_S$.

Since $\mathrm{Tr}_{E}(A_k)=0$, we have $C_S^{(0)}=0$. Therefore, $A_kH_S \otimes I_E=\sum_{i=1}^{d_E^2-1} (C_S^{(i)}H_S)\otimes G_E^{(i)}$, for which the partial trace over the environment vanishes.
 Similarly, we can show that   $\mathrm{Tr}_{E}(H_S \otimes I_E A_k)=0$, and so $\mathrm{Tr}_{E}([A_k,  H_S \otimes I_E])=0$.

To show that  $\mathrm{Tr}_{E}([A_k, I_S \otimes H_E])=0$, we first expand each $G_E^{(i)}$ in the basis of the eigenstates of $H_E$: 
\begin{equation}
\label{eq:12}
\begin{aligned}
G_E^{(i)}=\sum_{j,l=1}^{d_E} g_{jl}^{(i)} \vert j_{E}\rangle\langle l_{E}\vert,
\end{aligned}
\end{equation}
where $g_{jl}^{(i)}$ are complex coefficients.
Since $H_E\vert j_{E}\rangle =e_j\vert j_{E}\rangle$, with real eigenvalues $e_j$, we have 
\begin{equation}
\label{eq:13}
\begin{aligned}
G_E^{(i)}H_E-H_EG_E^{(i)}
=\sum_{j,l=1}^{d_E} g_{jl}^{(i)}(e_l-e_j) \vert j_{E}\rangle\langle l_{E}\vert,
\end{aligned}
\end{equation}
which is obviously traceless. So, for the commutator
\begin{equation}
\label{eq:14}
\begin{aligned}
\
[ A_k, I_S \otimes H_E]=[\sum_{i=1}^{d_E^2-1} C_S^{(i)}\otimes G_E^{(i)},  I_S \otimes H_E] \qquad \\
\qquad\qquad =\sum_{i} C_S^{(i)}\otimes (G_E^{(i)}H_E-H_EG_E^{(i)}), \
\end{aligned}
\end{equation}
the partial trace over the environment $E$ vanishes. $\qquad\blacksquare$

Form Lemma \ref{lem1} and Eq. \eqref{eq:9}, it is obvious that if we can show that $\mathrm{Tr}_{E}([A_k , H_{SE}])=0$, then $\mathrm{Tr}_{E}([A_k , H])=0$, and so, according to Eq.\eqref{eq:8},  $\mathrm{Tr}_{E}(A_{k+1})=0$.
Therefore, we can give the main result of this section, as the following:
\begin{propo}
\label{pro1}
 Consider two different initial states of the system-environment $\rho_{SE}$ and $\sigma_{SE}$ in Eq. \eqref{eq:5} such that their reduced states are the same, i.e., $\mathrm{Tr}_{E}(R)=0$.
 In addition, assume that the
   time evolution operator is given by  $U=\exp (-iHt)$. Expand the Hamiltonian $H$ as given in Eq. \eqref{eq:9}.
    Now, if we can show that $\mathrm{Tr}_{E}([A_k , H_{SE}])=0$, for arbitrary $A_k$ in Eq.\eqref{eq:8}, then, for both initial states $\rho_{SE}$ and $\sigma_{SE}$, the reduced dynamics of the system $S$ is the same for arbitrary $t$.
 \end{propo}

\section{Main result}  \label{sec: D}


As illustrated in Fig. \ref{Fig1},
 consider a spin chain from spin $1$ to spin $N$, such that they interact through the Hamiltonian given in Eq. \eqref{eq:b1}. We choose spins $1$ to $n-1$ as our system $S$, and spins $n$ to $N$ as the environment $E$. Comparing Eqs. \eqref{eq:b1} and \eqref{eq:9} shows that
\begin{equation}
\label{eq:15}
\begin{aligned}
H_{SE}=Z_{n-1} \otimes Z_n =Z_{n-1} \otimes Z_n \otimes I_{\Lambda\backslash (n-1,n)},
\end{aligned}
\end{equation} 
 where $I_{\Lambda\backslash (n-1,n)}$ denotes the identity operator on the whole  spin chain $\Lambda$, except spin $n-1$ and spin $n$.

First, note that the linear operator $A_k$, in Eq. \eqref{eq:8}, can be decomposed as
 \begin{equation}
\label{eq:16}
\begin{aligned}
A_k= C^{(0)}\otimes I_n+ C^{(1)}\otimes X_n + C^{(2)}\otimes Y_n + C^{(3)}\otimes Z_n,
\end{aligned}
\end{equation} 
 where $I_n$ is the identity operator on spin $n$. Also, $X_n$, $Y_n$ and $Z_n$ are the Pauli operators on spin $n$, and $C^{(\mu)}$ are linear operators on the other spins. 
 Assume that $\mathrm{Tr}_{\tilde{E}}(C^{(0)})=0$ and $\mathrm{Tr}_{\tilde{E}}(C^{(3)})=0$, where $\tilde{E}$ denotes all the spins in the environment $E$, except  spin $n$. Assuming that $\mathrm{Tr}_{\tilde{E}}(C^{(0)})=0$ results in  $\mathrm{Tr}_E (A_k)=0$.
In addition, assuming that $\mathrm{Tr}_{\tilde{E}}(C^{(3)})=0$, and using Eqs. \eqref{eq:15} and \eqref{eq:16}, it can be shown  simply that
 $\mathrm{Tr}_{E}([A_k , H_{SE}])=0$.
 So, as we have seen in the previous section, we have  $\mathrm{Tr}_{E}(A_{k+1})=0$.  
 
Let us emphasize  what is important for us is  that, for any $k$ in Eq. \eqref{eq:8}, $\mathrm{Tr}_E(A_k)$ vanishes. This leads to Eq. \eqref{eq:7} for arbitrary $t$,  and so the reduced dynamics of both initial states  $\rho_{SE}$ and $\sigma_{SE}$ is always the same.
   Using the decomposition of $A_k$ in Eq. \eqref{eq:16}, the requirement that $\mathrm{Tr}_E(A_k)=0$ means that we  must have  $\mathrm{Tr}_{\tilde{E}}(C^{(0)})=0$. Now, as we have seen in the previous paragraph, if for an $A_k$, we know that  $\mathrm{Tr}_{\tilde{E}}(C^{(0)})=0$ and $\mathrm{Tr}_{\tilde{E}}(C^{(3)})=0$, then we conclude that not only $\mathrm{Tr}_E(A_k)=0$, but also $\mathrm{Tr}_E(A_{k+1})=0$. So, adding the additional requirement $\mathrm{Tr}_{\tilde{E}}(C^{(3)})=0$ seems useful, as we will see in the  following Lemma:
  
 

 \begin{figure}
\includegraphics[width=8.75 cm]{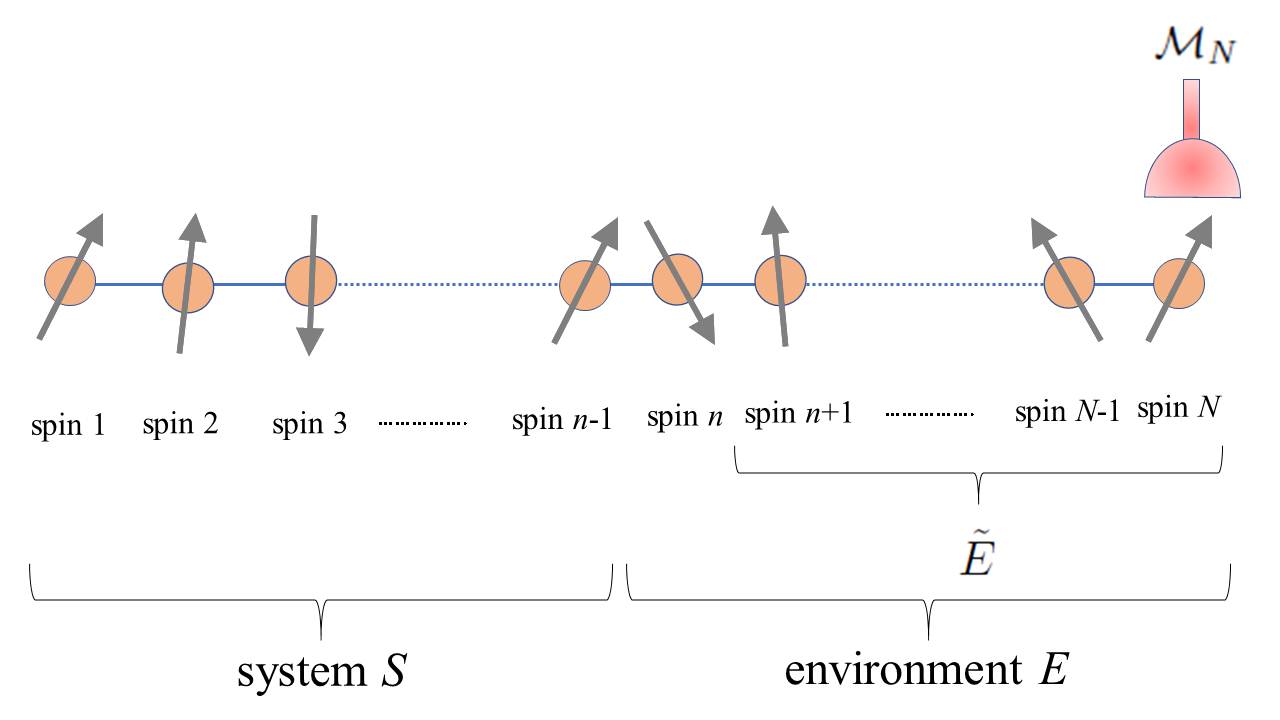}
\caption{Spin chain $\Lambda$, from spin $1$ to spin $N$. Each spin interacts with its nearest neighbors through the Hamiltonian  in Eq. \eqref{eq:b1}. The measurement $\mathcal{M}_N$ is performed on spin $N$, at $t=0$.
We divide the spin chain into two parts: from spin $1$ to spin $n-1$ as the system $S$, and from spin $n$ to spin $N$ as the environment $E$.
In addition, we divide the environment $E$ into two parts: spin $n$ and $\tilde{E}$.}
\label{Fig1}
\end{figure}

 \begin{lemma}
\label{lem 2}
If $A_k$, in Eq. \eqref{eq:8}, can be decomposed as Eq. \eqref{eq:16}, with $\mathrm{Tr}_{\tilde{E}}(C^{(0)})=0$ and $\mathrm{Tr}_{\tilde{E}}(C^{(3)})=0$, then   we have similarly
 \begin{equation}
\label{eq:17}
\begin{aligned}
A_{k+1}=  D^{(0)}\otimes I_n+ D^{(1)}\otimes X_n + D^{(2)}\otimes Y_n + D^{(3)}\otimes Z_n,
\end{aligned}
\end{equation} 
with $\mathrm{Tr}_{\tilde{E}}(D^{(0)})=0$ and $\mathrm{Tr}_{\tilde{E}}(D^{(3)})=0$.
\end{lemma}
\textit{Proof.}
 As we have seen after Eq. \eqref{eq:16}, when $\mathrm{Tr}_{\tilde{E}}(C^{(3)})=0$, then  $\mathrm{Tr}_{E}(A_{k+1})=0$. So, $\mathrm{Tr}_{\tilde{E}}(D^{(0)})=0$.
 
 In order to show that $\mathrm{Tr}_{\tilde{E}}(D^{(3)})=0$, we first decompose the Hamiltonian in Eq. \eqref{eq:b1}  as
\begin{equation}
\label{eq:18}
\begin{aligned}
H=Z_{n-1}\otimes Z_n + Z_n\otimes Z_{n+1}+I_n\otimes \tilde{H},
\end{aligned}
\end{equation}  
 where $\tilde{H}$ is a Hermitian operator on the other spins, except spin $n$. Using Eqs. \eqref{eq:16} and \eqref{eq:18}, and since $A_{k+1}=[A_k,H]$, we conclude that $D^{(3)}$ in Eq. \eqref{eq:17} is 
 \begin{equation}
\label{eq:19}
\begin{aligned}
D^{(3)}=[C^{(0)},  Z_{n-1}]+[C^{(0)},  Z_{n+1}]+[C^{(3)},  \tilde{H}].
\end{aligned}
\end{equation}
 As stated after Eq. \eqref{eq:16}, $C^{(0)}$ is a linear operator on $\cH_S \otimes \cH_{\tilde{E}}$, where $\cH_{\tilde{E}}$ denotes the Hilbert space of $\tilde{E}$, such that $\mathrm{Tr}_{\tilde{E}}(C^{(0)})=0$.
  In addition, $ Z_{n-1}$ and $ Z_{n+1}$ are Hermitian operators on $\cH_S \otimes \cH_{\tilde{E}}$ as $H_S^{\prime} \otimes I_{\tilde{E}}$ and $I_S \otimes H^{\prime}_{\tilde{E}}$, respectively, where $H_S^{\prime}$ is a Hermitian operator on $\cH_S$,  $I_{\tilde{E}}$ is the identity operator on $\cH_{\tilde{E}}$ and $H^{\prime}_{\tilde{E}}$ is a Hermitian operator on $\cH_{\tilde{E}}$.
 So, using Lemma \ref{lem1} (after replacing $E$ with $\tilde{E}$), we conclude that $\mathrm{Tr}_{\tilde{E}}([C^{(0)},  Z_{n-1}])=0$ and  $\mathrm{Tr}_{\tilde{E}}([C^{(0)},  Z_{n+1}])=0$.
 
Similarly, $C^{(3)}$ is a linear operator on $\cH_S \otimes \cH_{\tilde{E}}$ such that $\mathrm{Tr}_{\tilde{E}}(C^{(3)})=0$. In addition, $\tilde{H}$ is a  Hermitian operators on $\cH_S \otimes \cH_{\tilde{E}}$.
Since the Hamiltonian in Eq. \eqref{eq:b1} only includes the nearest  neighbors interaction, $\tilde{H}$ includes terms as $H_S^{\prime} \otimes I_{\tilde{E}}$ or $I_S \otimes H^{\prime}_{\tilde{E}}$. So, again using Lemma \ref{lem1}, we conclude that  $\mathrm{Tr}_{\tilde{E}}([C^{(3)},  \tilde{H}])=0$. Therefore, we have shown that $\mathrm{Tr}_{\tilde{E}}(D^{(3)})=0$. $\ \ \qquad\qquad\qquad\quad\qquad\qquad\blacksquare$

Let us decompose the linear operator $R=A_0$, in Eq. \eqref{eq:5}, as Eq. \eqref{eq:16}:
\begin{equation}
\label{eq:20}
\begin{aligned}
R= R^{(0)}\otimes I_n+ R^{(1)}\otimes X_n + R^{(2)}\otimes Y_n + R^{(3)}\otimes Z_n.
\end{aligned}
\end{equation} 
Now, since in Eq. \eqref{eq:5} we have chosen $\rho_{SE}$ and $\sigma_{SE}$ such that their reduced states of the system $S$ are the same, we have $\mathrm{Tr}_{\tilde{E}}(R^{(0)})=0$.
Lemma \ref{lem 2} shows that if  we choose $R$ appropriately, i.e., if in addition $\mathrm{Tr}_{\tilde{E}}(R^{(3)})=0$, then, for arbitrary $k$ in Eq. \eqref{eq:16}, we have $\mathrm{Tr}_{\tilde{E}}(C^{(0)})=0$. Therefore, $\mathrm{Tr}_{E}(A_k)=0$, for all $k$ in Eq. \eqref{eq:8}, and so, for all the times $t$, the reduced dynamics of the system $S$ remains the same for both initial states of the system-environment $\rho_{SE}$ and $\sigma_{SE}$ in Eq. \eqref{eq:5}.
 
We choose the initial state of the system-environment at $t=0$ as  
\begin{equation}
\label{eq:21}
\begin{aligned}
\rho_{SE}=\rho_{S\tilde{E}}\otimes \rho_{n},
\end{aligned}
\end{equation}  
 where $\rho_{S\tilde{E}}$ is an arbitrary state on $\cH_S \otimes \cH_{\tilde{E}}$, and $\rho_{n}$ is an arbitrary state of spin $n$, in the $xy$ plane of the Bloch sphere \cite{1}:
 \begin{equation}
\label{eq:22}
\begin{aligned}
\rho_{n}=\frac{1}{2}(I_n+r_x X_n +r_y Y_n),
\end{aligned}
\end{equation} 
 where $r_x$ and $r_y$  are real coefficients. So, $\rho_{SE}$ in Eq. \eqref{eq:21} can be decomposed as
 \begin{equation}
\label{eq:23}
\begin{aligned}
\rho_{SE}=\bar{R}^{(0)}\otimes I_n+ \bar{R}^{(1)}\otimes X_n + \bar{R}^{(2)}\otimes Y_n,
\end{aligned}
\end{equation} 
 where $\bar{R}^{(\mu)}$ are Hermitian operators on $\cH_S \otimes \cH_{\tilde{E}}$.
 
If at time $t=0$ we perform an instantaneous measurement $\mathcal{M}_N$ on spin $N$, then the whole state of the system-environment, at the same time $t=0$, changes to
 \begin{equation}
\label{eq:24}
\begin{aligned}
\sigma_{SE}=\hat{R}^{(0)}\otimes I_n+ \hat{R}^{(1)}\otimes X_n + \hat{R}^{(2)}\otimes Y_n \\
=\rho_{SE} +R, \qquad\qquad\qquad\qquad\qquad\quad \
\end{aligned}
\end{equation} 
 where $\hat{R}^{(\mu)}=\mathrm{id}_{\Lambda\backslash (nN)}\otimes\mathcal{M}_N (\bar{R}^{(\mu)})$.
 The map $\mathcal{M}_N$ is a trace-preserving (completely positive) map on spin $N$, and  
 $\mathrm{id}_{\Lambda\backslash (nN)}$ is the identity map on the whole spin chain $\Lambda$, except spins $n$ and $N$. 
 
 Now, if we decompose $R$ in Eq. \eqref{eq:24} as  Eq. \eqref{eq:20}, we have
\begin{equation}
\label{eq:25}
\begin{aligned}
R^{(\mu)}=\hat{R}^{(\mu)}-\bar{R}^{(\mu)}.
\end{aligned}
\end{equation} 
Since $\mathcal{M}_N$ is  trace-preserving, we have $\mathrm{Tr}_{\tilde{E}}(R^{(0)})=0$.
In other words, performing the measurement $\mathcal{M}_N$ on spin $N$, which is a part of the environment $E$, does not affect (the reduced state of) the system $S$, and so $\mathrm{Tr}_{E}(R)=0$, which results in $\mathrm{Tr}_{\tilde{E}}(R^{(0)})=0$.

In addition, we have chosen $\rho_{SE}$ in Eq. \eqref{eq:23} such that $\bar{R}^{(3)}=0$. So, $\hat{R}^{(3)}$ in Eq. \eqref{eq:24} and $R^{(3)}$ in Eq. \eqref{eq:25} are also zero.
Therefore, the operator $R=A_0$ in Eq. \eqref{eq:24} has all the requirements needed in Lemma \ref{lem 2}. Consequently, $\mathrm{Tr}_{E}(A_k)=0$, for all $k$ in Eq. \eqref{eq:8}, and so the reduced state of the system $S$, for both initial states $\rho_{SE}$ in Eq. \eqref{eq:23} and $\sigma_{SE}$ in Eq. \eqref{eq:24},  remains the same for all the times $t$.

In other words, whether or not we perform the measurement $\mathcal{M}_N$ on spin $N$ at $t=0$, and so whether the initial state of the whole spin chain at $t=0$  is $\sigma_{SE}$ in Eq. \eqref{eq:24} or  $\rho_{SE}$ in Eq. \eqref{eq:23},
 the reduced dynamics of the system $S$ (and so spin $1$ as a part of $S$) remains the same for all $t \geq 0$. Therefore, spin $1$ will never become aware of performing the measurement $\mathcal{M}_N$ on spin $N$.

Let us summarize our main result of this paper, in the following Proposition:
\begin{propo}
\label{pro2}
 Consider a spin chain, from spin $1$ to spin $N$, which interact with each other through the Hamiltonian $H=\sum_{j=1}^{N-1} Z_j \otimes Z_{j+1}$. We divide this spin chain into two parts: We consider spins $1$ to $n-1$ as the system $S$, and spins $n$ to $N$ as the environment $E$. Next, we choose the initial state of the whole spin chain at $t=0$ as $\rho_{SE}$  in Eq. \eqref{eq:23}. Performing an instantaneous trace-preserving measurement  $\mathcal{M}_N$ on spin $N$ changes  the initial state of spin chain to $\sigma_{SE}$ in Eq. \eqref{eq:24}, at the same time $t=0$. But, the reduced state of the system $S$  remains unchanged at this initial instant $t=0$, i.e., $\mathrm{Tr}_E(\sigma_{SE})= \mathrm{Tr}_E(\rho_{SE})$. Using Lemma \ref{lem 2}, we  showed that Eq. \eqref{eq:7} is satisfied, and so the reduced state of the system $S$,  for all the times, remains the same, for both initial states $\sigma_{SE}$ and $\rho_{SE}$. In other words, performing or not performing the measurement  $\mathcal{M}_N$ on spin $N$ at $t=0$ does not affect the reduced dynamics of the system $S$ for any $t \geq 0$. Therefore, any spin in the system $S$, including spin $1$, will never become aware of performing the measurement  $\mathcal{M}_N$ on spin $N$ at $t=0$.
 \end{propo}

\begin{rem}
\label{rem1}
The minimum $N$ for which the results of this section can be applied is $N=3$. Then, spin $1$ is the system $S$, $n=2$ and spin $3$ is the subspace $\tilde{E}$.
\end{rem}

\begin{rem}
\label{rem2}
In order to achieve Proposition \ref{pro2}, we only require that initial $\rho_{SE}$ be as Eq. \eqref{eq:23}.  The state given in  Eq. \eqref{eq:21} is only an example of such $\rho_{SE}$.
In fact, any initial state as $\rho_{SE}=\bar{R}^{(0)}\otimes I_n+ \bar{R}^{(1)}\otimes X_n + \bar{R}^{(2)}\otimes Y_n +\bar{R}^{(3)}\otimes Z_n$, with  $\mathrm{Tr}_{\tilde{E}}(\bar{R}^{(3)})=0$, is also appropriate.
\end{rem}

In Sec. \ref{sec: B2}, we have seen that any instantaneous measurement $\mathcal{M}_N$ on spin $N$ can be modeled as performing a unitary operator $U_{NP}(\theta)=\exp(-i\theta \tilde{H}_{NP})$, and then tracing over the probe $P$. Therefore, the final state of the whole spin chain $\Lambda$, after the unitary time evolution $U_\Lambda =U=\exp(-iHt)$ with $H$ in Eq. \eqref{eq:b1}, is given by Eq. \eqref{eq:b21}.
So, the final state of the system $S$ is 
\begin{equation}
\label{eq:25a}
\begin{aligned}
\sigma^{(\theta)}_S(t)=\mathrm{Tr}_{E}(\sigma^{(\theta)}_\Lambda (t)) \qquad\qquad\qquad \qquad\qquad\quad    \\
=\mathrm{Tr}_{EP} \circ 
\mathcal{U}(t) \circ \mathcal{U}_{NP}(\theta)
  [\rho_\Lambda(0)\otimes \rho_P(0)], 
\end{aligned}
\end{equation} 
which is, in general, a function of the parameter $\theta$.
Assuming that  $\lbrace E_\gamma \rbrace$ is a POVM on the whole system $S$ (not only on  spin $1$), the probability of the outcome $\gamma$ is given by $P(\gamma \vert \theta)=\mathrm{Tr}[E_\gamma \sigma_S^{(\theta)} (t)]$, and the Fisher information of this POVM is given by  Eq. \eqref{eq:b26}.

On the other hand, not performing the  measurement $\mathcal{M}_N$ on spin $N$,  the final state of the system $S$, using Eq. \eqref{eq:b22}, is
\begin{equation}
\label{eq:25b}
\begin{aligned}
\rho_S(t)=\mathrm{Tr}_{E}(\rho_\Lambda (t)) 
=\mathrm{Tr}_{EP} \circ 
\mathcal{U}(t) 
  [\rho_\Lambda(0)\otimes \rho_P(0)].
\end{aligned}
\end{equation}  
 When the initial state of the spin chain $\rho_\Lambda(0)$ is given by Eq. \eqref{eq:23}, we have seen that $\sigma^{(\theta)}_S(t)=\rho_S(t)$, for arbitrary $t$. Therefore,  $\sigma^{(\theta)}_S(t)$ does not depend on $\theta$, and the Fisher information in   Eq. \eqref{eq:b26} is zero, for arbitrary POVM $\lbrace E_\gamma \rbrace$. In other words, when $\rho_\Lambda(0)$ is given by Eq. \eqref{eq:23}, no measurement on the system $S$ is sensitive to $\theta$.

We end this section with the following example. Consider the case that $N=3$. Assume that the initial state of the spin chain is
 \begin{equation}
\label{eq:25c}
\begin{aligned}
\vert \psi_\Lambda (0) \rangle =
\vert \Psi_{13} \rangle \vert (X+)_2 \rangle      \qquad\qquad\qquad   \qquad\qquad\qquad  \qquad  \  \\
 =\frac{1}{\sqrt{2}} \left(\vert (Z+)_1 (Z+)_3 \rangle + \vert (Z-)_1 (Z-)_3 \rangle \right)  \vert (X+)_2 \rangle ,  
\end{aligned}
\end{equation}  
 where $\vert (Z\pm)_i \rangle$ are the eigenstates of $Z_i$ with  eigenvalues $\pm 1$, and  
 $\vert (X+)_2 \rangle$ is the eigenstate of $X_2$ with  eigenvalue $+1$.
 This initial state is as  Eq. \eqref{eq:21}. So, according to  Proposition \ref{pro2}, $\sigma_1 (t)$, i.e., the reduced state of spin $1$ at arbitrary time $t$, is not affected by any measurement $\mathcal{M}_3$ performed on spin $3$ at $t=0$.
 In other words, although the initial state of spins $1$ and $3$ is the maximally entangled state $\vert \Psi_{13} \rangle $, and also all the spins interact through the Hamiltonian in Eq. \eqref{eq:b1}, spin $1$ never become aware of the measurement $\mathcal{M}_3$ performed at $t=0$ on spin $3$ .
 
Let us emphasize  when we state that spin $1$ never become aware of the measurement $\mathcal{M}_3$ performed  on spin $3$, we mean that the reduced state  $\sigma_1 (t)$ does not change, performing or not performing $\mathcal{M}_3$. Therefore, having access only to spin $1$, no (expectation value of any) measurement on spin $1$ is affected by $\mathcal{M}_3$.
 This is the case even though, for example, measuring spin $3$ in the $z$-direction changes the initial state of the whole spin chain in Eq. \eqref{eq:25c}, and so its consequent evolution governed by the  Hamiltonian in Eq. \eqref{eq:b1}, dramatically.

One may step beyond and argue that
 the whole information in spin $1$ is what  can be extracted from its reduced state. Then, the above result can be interpreted as the following: No information of performing $\mathcal{M}_3$  on spin $3$ will achieve spin $1$.

It is worth noting that such kind of assumption  is, in fact,  the basis of the Deutsch–Hayden approach to  quantum mechanics \cite{2}. They have assumed that the information is local. So, 
 the information in a part of a composite system is what that can be extracted from its reduced state.


\section{Non-instantaneous measurement} \label{sec: E}

In the previous section, we  showed that it is possible to find cases for which  performing an instantaneous measurement on spin $N$ at $t=0$ does not have any influence on the  reduced state of spin $1$ for all $t \geq 0$.

Now, one may ask whether instantaneous  measurement is possible or not.  
According to the text-book quantum mechanics, instantaneous  measurement seems necessary: Consider a particle with a spatially widespread wave function, which is detected in a localized detector.
 It seems that we should accept that its wave function  collapses instantaneously, from the pre-measurement  widespread one to a localized one, which is confined within the detector.  
 
Nevertheless, in many cases, one can model the measurement procedure as the following \cite{15}.  Step 1: The quantum system $Q$, which we want to measure the observable $\mathcal{M}_{Q}$  (with projectors $\Pi_l$)  on it, interacts with another system, the probe $P$, such that the whole state of the system-probe $\rho_{QP}$ becomes correlated. Step 2: A projective measurement is performed on $P$, in a special basis, called the pointer states ${\vert l_P \rangle }$. Now, if the measurement result is ${\vert l_P \rangle }$, the state of the system  $Q$ is left within the subspace spanned by $\Pi_l$.
 
 Step 1 is deterministic. It can be modeled  simply assuming that the system $Q$ and the probe $P$ interact through a unitary operator, for a period of time $\delta$, such that the initial uncorrelated state of the system-probe changes to an entangled (a correlated) one \cite{13, 9, 11}. One can also consider more involved models, such as considering an environment $\mathcal{E}$  which interacts with $P$ \cite{14, 8}, or averaging over different periods $\delta$ \cite{12}. 
 
Even if we assume that the probabilistic step 2 is performed instantaneously, as predicted by the standard text-book quantum mechanics, step 1 obviously takes some time. 
Consider the case that the initial state of the system-probe, before interacting with each other, is $\vert \psi_Q \rangle \otimes  \vert 0_P \rangle$. An estimate for the time needed to evolve from this initial state to the entangled state $\vert \phi_{QP} \rangle =\sum_l \Pi_l \vert \psi_Q \rangle \otimes \vert l_P \rangle$ is given in Ref. \cite{9}. Also, in Ref. \cite{8}, a lower bound on the time needed to go from the pure state $\vert \phi_{QP} \rangle$ to the mixed state $\rho_{QP}=\sum_l \Pi_l \vert \psi_Q \rangle \langle \psi_Q \vert \Pi_l \otimes \vert l_P \rangle \langle l_P \vert$ is presented.
 One may argue that the time needed to go from $\vert \phi_{QP} \rangle$ to  $\rho_{QP}$, as a result of the interaction between $P$ and $\mathcal{E}$, can be considered as the time that the collapse of the wave function $\vert \phi_{QP} \rangle$  lasts \cite{14, 15, 8}. But, we think that the collapse is a probabilistic event, occurred only through step 2.

The question arose as to 
whether our results in the previous section can be applied to the case that
 the measurement procedure on spin $N$ includes the  two above mentioned steps, and so takes some time $\delta $ for the probe $P$ to interacts with spin $N$.

 Assume that the  measurement  $\mathcal{M}_N$ on spin $N$ lasts from $t=0$ to $t=\delta$.  Denote the initial state of the whole spin chain at $t=0$ as $\rho_{\Lambda}(0)$. So, $\rho_{\Lambda}(\delta)=\exp(-iH\delta)\rho_{\Lambda}(0)\exp(iH\delta)$ gives the state of the spin chain, after the time interval $\delta$, if the  measurement  $\mathcal{M}_N$ were not performed.
In addition, denote   the state of the spin chain at $t=\delta$, right after the measurement  $\mathcal{M}_N$, as $\sigma_{\Lambda}(\delta)$. Now, if we have  $\sigma_{\Lambda}(\delta)= \mathrm{id}_{\Lambda\backslash N}\otimes\mathcal{M}_N [\rho_{\Lambda}(\delta)]$, where $\mathcal{M}_N$ is a linear trace-preserving map on spin $N$, and $\mathrm{id}_{\Lambda\backslash N}$ is the identity map on the rest of the spin chain, then we can follow a similar line of reasoning, as given from Eq. \eqref{eq:23} onward, to show that, choosing $\rho_{\Lambda}(\delta)$ appropriately, spin $1$ will never become aware of performing $\mathcal{M}_N$ on spin $N$.

 Therefore, at least,  when 
 $\delta$  is sufficiently small such that $U=\exp(-iH\delta)$ can be approximated by the identity operator, our results in the previous section can be used approximately, since $\rho_{\Lambda}(\delta)$ is almost the same as  $\rho_{\Lambda}(0)$, and $\sigma_{\Lambda}(\delta)$ is approximately given by $\mathrm{id}_{\Lambda\backslash N}\otimes\mathcal{M}_N [\rho_{\Lambda}(0)]$.

More precisely, decompose $H$ in Eq. \eqref{eq:b15} into two parts: $H=H_{com}+H_{ncom}$, where $H_{com}$ commutes with $H_{NP}(t)$, but $H_{ncom}$ does not commute with $H_{NP}(t)$. Also, remember  we have assumed that $H_{NP}(t)$ is turned on during the time interval $[t_1, t_2]$. Here, we have chosen $t_1=0$ and $t_2=\delta$. Now, consider the case that $\vert\vert H_{com} \vert\vert \gg \vert\vert H_{ncom} \vert\vert$. This is the case when $H$ is as Eq. \eqref{eq:b1}, and the spin chain  is large enough. If, in addition, we have $ \vert\vert H_{NP}(t) \vert\vert \gg \vert\vert H_{ncom} \vert\vert$ 
 during the time interval $[0, \delta]$, we can neglect the effect of $H_{ncom}$  during this short time interval. So, we can follow a similar line of reasoning as what led to Eqs. \eqref{eq:b11}-\eqref{eq:b13}. Consequently, we  have
\begin{equation}
\label{eq:25d}
\begin{aligned}
\sigma_\Lambda (\delta)\approx
\mathrm{Tr}_{P} \circ 
\mathcal{U}_{com} \circ \mathcal{\tilde{U}}_{NP} (\delta)
  [\rho_\Lambda(0)\otimes \rho_P(0)]    \\
  =\mathrm{Tr}_{P} \circ \mathcal{\tilde{U}}_{NP} (\delta) \circ \mathcal{U}_{com}  [\rho_\Lambda(0)\otimes \rho_P(0)]    \\
  =(\mathrm{id}_{\Lambda\backslash N}\otimes\mathcal{M}_N )\circ \mathcal{U}_{com}  [\rho_\Lambda(0)], \qquad    \quad
\end{aligned}
\end{equation}
where $\rho_P(0)$ is the initial state of the probe $P$ and $U_{com}=\exp(-iH_{com}\delta)$.
Also,  $\tilde{U}_{NP}$ is the unitary evolution operator generated by $H_{NP}(t)$ from $t=0$ to $t=\delta$.
In addition, $\mathcal{M}_N [*]=\mathrm{Tr}_{P} \circ \mathcal{\tilde{U}}_{NP} (\delta)  [ *\otimes \rho_P(0)]$ is a completely positive trace-preserving map which describes the measurement on spin $N$.

On the other hand, not performing the measurement $\mathcal{M}_N$ during the time interval $[0, \delta]$, we have
\begin{equation}
\label{eq:25e}
\begin{aligned}
\rho_\Lambda (\delta)= \mathcal{U} [\rho_\Lambda(0)] \approx
\mathcal{U}_{com}
  [\rho_\Lambda(0)],
\end{aligned}
\end{equation}
where $U=\exp(-iH\delta)$.
Comparing Eqs. \eqref{eq:25d} and \eqref{eq:25e}, we conclude that 
\begin{equation}
\label{eq:25f}
\begin{aligned}
\sigma_{\Lambda}(\delta)\approx \mathrm{id}_{\Lambda\backslash N}\otimes\mathcal{M}_N [\rho_{\Lambda}(\delta)].
\end{aligned}
\end{equation}
 So,  choosing $\rho_{\Lambda}(\delta)$  as Eq.  \eqref{eq:23}, performing $\mathcal{M}_N$ on spin $N$ influences neglectably  the reduced state of 
 spin $1$  for all $t \geq \delta$.

\section{Summary} \label{sec: F}


In this paper, we considered a spin chain, from spin $1$ to spin $N$, which interact through the free Hamiltonian $H$ in Eq. \eqref{eq:b1}. Performing a measurement on spin $N$, we studied whether its influence will be achieved spin $1$ or not

Any arbitrary measurement (operation) on spin $N$, regardless of whether it is instantaneous or not, 
can be modeled introducing an interaction Hamiltonian $H_{NP}$ between spin $N$ and the probe $P$.

In Sec.  \ref{sec: B1}, we considered the case that $[H, H_{NP}]=0$. We saw that other spins (including spin $1$) never become aware of performing such measurement on spin $N$. In addition, as we have seen in Eqs. \eqref{eq:b12} and \eqref{eq:b13}, such measurements can be considered instantaneous, even if they last a period of time.

Then, in Sec.  \ref{sec: B2}, we considered the case that $[H, H_{NP}]\neq 0$. Using the  Lieb-Robinson bound in Eq. \eqref{eq:b17}, we showed that, regardless of whether the measurement on spin $N$ is instantaneous or not, we expect that its effect  achieves spin $1$ after some while, in general.

The  Lieb-Robinson bound gives us an upper bound on the speed of  propagation of  information in the spin chain. So, we asked, when $[H, H_{NP}]\neq 0$,  whether it is possible to find cases for which the optimal lower bound, i.e., the zero speed, occur.

In Sec.  \ref{sec: D}, we  showed that if we choose the initial state of the spin chain as Eq. \eqref{eq:23}, then the reduced state of spin $1$ will be never affected by an instantaneous  measurement $\mathcal{M}_N$ performed on spin $N$ at $t=0$. This is the case although $[H, H_{NP}]\neq 0$, and
even if the initial $\rho_{S\tilde{E}}$ in Eq. \eqref{eq:21} is correlated. (Equation  \eqref{eq:21} is a 
 special case of Eq. \eqref{eq:23}.)

The measurement procedure, in general, includes some steps which take some time and so the measurement  is not instantaneous. 
 Therefore,  in Sec.  \ref{sec: E}, we discussed whether our results in Sec.  \ref{sec: D} can be generalized to  non-instantaneous measurements too.  

Remembering this result of Sec.  \ref{sec: B1} that when $[H, H_{NP}]=0$ the measurement can be considered instantaneous, we decomposed $H$ as $H=H_{com}+H_{ncom}$, where $[H_{com}, H_{NP}]=0$ but $[H_{ncom}, H_{NP}]\neq 0$. Then, we considered the case   that, during the short time interval of the measurement $\delta$,  the effect of  $H_{ncom}$, compared to the effects of $H_{com}$ and $H_{NP}$, is neglectable.
For this case, we saw that Proposition \ref{pro2} remains approximately valid, shifting the initial time from $t=0$ to $t=\delta$.


%
%
%
%


\begin{thebibliography}{1}
\bibitem{b4} The norm of an operator $O$ is defined as $\vert\vert O\vert\vert = \max_{\vert\psi\rangle}\vert O \vert\psi\rangle\vert  $, where the maximum is taken over all normalized kets $\vert \psi \rangle$. So, when $O$ is Hermitian, $\vert\vert O\vert\vert$ is equal to  the absolute value of its largest eigenvalue.
\end{thebibliography}
\end{document}